# Production of Higgs Boson in ultra-peripheral heavy ion collisions with two-photon processes

Gongming Yu[1], Wenlong Sun[2]

[1]College of Physics and Technology, Kunming University, Kunming 650214, China
[2]Fundamental Science on Nuclear Safety and Simulation Technology Laboratory, Harbin Engineering University, Harbin 150000, China
Email: ygmanan@kmu.edu.cn, 18109133295@163.com

We calculated the production of the Higgs boson (H) by two-photon interaction with the equivalent photon approximation in nucleus-nucleus collision, proton-nucleus collision, and proton-proton collision. The numerical results show that the experimental study of the Higgs boson in ultra-peripheral collisions is feasible at the energies of the relativistic heavy ion collider (RHIC) and the large hadron collider (LHC).

## I. INTRODUCTION

The Standard Model of particle physics describes the known elementary particles and forces (except gravity) that make up our universe. One of the central features of the Standard Model is that there are fields that interact with elementary particles. The quantum excitation of this feld, known as the Higgs feld, manifests itself as the Higgs boson, the only fundamental particle with no spin[1,2].

Theoretical studies of the two-photon mechanism could date back to 1934, Williams, Landau and Lifshitzo investigated the generation of electron-positron pairs through this mechanism. Around 1960, when the Colliding Beam Facility became operational, the "two-photon mechanism" attracted some attention. Calogero and Zemach studied the two-photon production process for neutron pairs in electron-electron collision [3]. Limited by the conditions at the time, the process of two-photon interaction to produce Higgs boson could not be realized. But with the first collision at Brookhaven National Laboratory's (BNL) Relativistic Heavy Ion Collider in June 2000, heavy ion physics entered a new phase. Interactions of atomic nuclei at such high energies have so far only been observed in cosmic ray interactions. Many interesting physics topics can now be studied in the laboratory[4]. The two-photon mechanism has once again attracted attention.

In the standard model, the Higgs mechanism is used to break the electroweak symmetry, which indicates the existence of the Higgs boson[5]. About 60 years ago, theoretical physicists proposed that a field permeates the universe and provides energy for a vacuum. This field explains why elementary particles have mass[6]. In recent decades, a large number of precise measurements have provided indirect support for the existence of this field, but due to various factors, the existence of the Higgs boson has not been confirmed.

On 4 July 2012, scientists and journalists gathered at CERN, and remotely around the world, for the announcement of the discovery of a new fundamental particle, the Higgs boson[7]. The discovery, by the ATLAS and CMS collaborations at the Large Hadron Collider (LHC)[8], came almost 50 years after theorists had postulated the existence of such a particle[9].

Others have done a lot of work on collisions that create Higgs bosons. Georgi et al studied the annihilation of digluons in proton-proton collisions[10],Higgs boson production with large transverse momentum in hadron collisions was studied by Baur et al.[11],Aad et al. studied the Higgs boson produced by the two-photon interaction of proton-proton collisions with center-of-mass energies of 7 TeV and 8 TeV[12].

Bertulani et al. study electromagnetic processes in relativistic heavy ion collisions[13], Contreras et al.studied ultraperipheral heavy ion collisions at the LHC [14],ultraperipheral nuclear collisions were studied by Klein et al. have provided assistance for this research[15].

The ultraperipheral collision (UPC) of heavy ions and protons is the energy front of electromagnetic interactions. Two-photon collisions are studied at much higher collision energies than elsewhere[16].In this paper, we will study the production of the Higgs boson produced by the collision of two virtual photons emitted by two charged particles in the case of ultra-peripheral collisions (the collision parameter $b > R_A + R_B$, where $R_A$ and $R_B$ are the nuclear radius of the collision particles A and B, respectively).

In this paper, we report the feasibility study of two-photon interaction at LHC and RHIC. The process of Higgs boson production by two-photon interaction in LHC and RHIC energy regions were presented in the Sec.II. The numerical values results of the Higgs boson produced by two-photon interactions for lead-lead, lead-proton and proton-proton collisions at LHC energies and gold-gold and uranium-uranium, gold-proton and proton-proton collisions at RHIC energies are plotted in Sec.III. The conclusion is given in Sec.IV.

## II. GENERAL FORMALISM

Photons emitted by high-speed nuclei can be regarded as real photons, because their virtual degree is less than $(\hbar c/R_A)^2$, which can be ignored [17]. In the equivalent photon approximation, the invariant matrix element square is further factorized as the convolution of the independent two photon spectral functions and the invariant matrix element square of the double real photon interaction process .

The cross section of the Higgs particle produced by the photon process can also be decomposed approximately into the elementary cross section of γγ→H and the γγ luminosity according to the equivalent photon, and the cross section of the final state with energy W can be written as[18]

$$d\sigma = \hat{\sigma}_{\gamma\gamma \to H}(W) dN_1(\omega_1, q_1^2) dN_2(\omega_2, q_2^2)$$
$$= d\omega_1 d\omega_2 \hat{\sigma}_{\gamma\gamma \to H}(W) \frac{dN_1(\omega_1, q_1^2)}{d\omega_1} \frac{dN_2(\omega_2, q_2^2)}{d\omega_2}, \quad (1)$$

where the energies of the photons emitted from the nucleus are $\omega_{1,2} = \frac{W}{2} \exp(\pm y)$, with $W^2 = 4\omega_1\omega_2$, and the transformations $d\omega_1 d\omega_2 = (W/2) dW dy$ can be performed. Then, the differential cross section of the Higgs boson produced by the two-photon interaction can be written as

$$d\sigma_{H_0} = \hat{\sigma}_{\gamma\gamma \to H_0}(W) dN_1(\omega_1, q_1^2) dN_2(\omega_2, q_2^2)$$
$$= d\omega_1 d\omega_2 \hat{\sigma}_{\gamma\gamma \to H_0}(W) \frac{dN_1(\omega_1, q_1^2)}{d\omega_1} \frac{dN_2(\omega_2, q_2^2)}{d\omega_2}, \quad (2)$$

Utilizing the decay width of the known $n_1 S_0$ state and the sum of all state, the total cross-section of the real photogenic process produced by the exotic electromagnetic confinement system can be easily written as[15]

$$\hat{\sigma}_{\gamma\gamma \to H^0}(W) = 8\pi^2 (2J+1) \frac{\Gamma_{H^0 \to \gamma\gamma}}{M} \delta(W^2 - M^2), \quad (3)$$

where J and M are the spin and mass of the produced Higgs boson, respectively.

In the equivalent photon approximate, that is similar to a real photon, the nuclear photon spectrum function can be written as[19]

$$\frac{dN_N(\omega, q^2)}{d\omega} = \frac{Z^2 \alpha}{\pi \omega} \int d^2 q_T \frac{q_T^2}{\left(q_T^2 + \frac{\omega^2}{\gamma^2}\right)^2} F_N^2(q^2), \quad (4)$$

where $q^2 = \left(q_T^2 + \omega^2/\gamma^2\right)^2$ the 4-momentum transfer of the relativistic nuclei projectile, and $F_N(q^2)$ is the nuclear form factor of the equivalent photon source.

The photon spectrum function of the proton can be written as[20-26]

$$\frac{dN_P(\omega, q^2)}{d\omega} = \frac{\alpha}{\pi \omega} \int d^2 q_T \frac{q_T^2}{\left(q_T^2 + \frac{\omega^2}{\gamma^2}\right)^2} F_P^2(q^2), \quad (5)$$

In the calculation, we use the simple approximation of the nuclear form factor. This approximation corresponds to the decreased charge distribution of the index of the nucleus, that is, the adjustment of the average square radius to adapt to the experimental value. In the case of the nucleus, the single-pole shape factors are given by

$$F_N(q) = \frac{\Lambda^2}{\Lambda^2 + q^2}, \quad (6)$$

For protons, the form factor is usually assumed as

$$F_P(q) = \frac{1}{\left(1 + \frac{q^2}{0.71 GeV^2}\right)^2}, \quad (7)$$

Here, it is known that the $\Lambda = 0.091 \text{GeV}$ of $^{197}\text{Au}$, the $\Lambda = 0.088 \text{GeV}$ of $^{208}\text{Pb}$, and the $\Lambda = 0.065 \text{GeV}$ of $^{238}\text{U}$. And $M_{H^0} = 124 \text{GeV}$.

In the semi-coherent two-photon interaction process at the ultra-peripheral nucleus-nucleus collisions, the momentum for photons are $q_1 = (\omega_1, q_{1T}, q_{1z})$ and $q_2 = (\omega_2, q_{2T}, q_{2z})$, the total transverse momentum is $p_T = q_{1T} + q_{2T} \approx q_{1T}$, where $q_{iT}$ is the transverse momentum of the $i$-th photon. When high-energy heavy ions collide, the momentum satisfy $q_{1z} \sim 0, \omega_1 \sim |\vec{q}_{1T}| \sim q_{1T}$, $d^2 p_T = 2\pi p_T dp_T = \pi dp_T^2$. Therefore, the differential section of the Higgs Boson ($H^0$) of the two-photon interaction during the inclusive process of ultraperipheral nuclear-nuclear collision can be expressed as

$$\frac{d\sigma_{AB \to AH^0B}}{d^2p_T dy} = \frac{Z_1^2 Z_2^2 \alpha^2}{\pi^3} 8\pi^2 (2J+1) \frac{\Gamma_{H^0 \to \gamma\gamma}}{M_{H^0}^3} \frac{P_T^2}{\left(P_T^2 + \frac{\omega_1^2}{\gamma^2}\right)^2} \left[F_N\left(P_T^2 + \frac{\omega_1^2}{\gamma^2}\right)\right]^2$$

$$\times \int d^2 q_{2T} \frac{q_{2T}^2}{\left(q_{2T}^2 + \frac{\omega_2^2}{\gamma^2}\right)^2} [F_N(q_{2T}^2 + \frac{\omega_2^2}{\gamma^2})]^2, \tag{8}$$

where $\gamma$ is the relativistic factor, $q_{iT}$ is the transverse momentum of the i-th photon, and the transverse momentum of photon is $q_{2T} > 0.2$GeV due to the single track acceptance condition .

In the inclusive process, the differential section of Higgs Boson ( $H^0$ ) of the two-photon interaction of ultraperipheral proton-nuclear collision can be expressed as

$$\frac{d\sigma_{pA \to pH^0A}}{d^2p_T dy} = \frac{Z^2 \alpha^2}{\pi^3} 8\pi^2 (2J+1) \frac{\Gamma_{H^0 \to \gamma\gamma}}{M_{H^0}^3} \frac{p_T^2}{\left(p_T^2 + \frac{\omega_1^2}{\gamma^2}\right)^2}$$

$$\times \left\{ \begin{array}{l} \left[F_N\left(P_T^2 + \frac{\omega_1^2}{\gamma^2}\right)\right]^2 \int d^2 q_{2T} q_{2T}^2 \frac{[F_P(q_{2T}^2 + \frac{\omega_2^2}{\gamma^2})]^2}{\left(q_{2T}^2 + \frac{\omega_2^2}{\gamma^2}\right)^2} \\ + \left[F_P\left(P_T^2 + \frac{\omega_1^2}{\gamma^2}\right)\right]^2 \int d^2 q_{2T} q_{2T}^2 \frac{[F_N(q_{2T}^2 + \frac{\omega_2^2}{\gamma^2})]^2}{\left(q_{2T}^2 + \frac{\omega_2^2}{\gamma^2}\right)^2} \end{array} \right\},$$

$$\tag{9}$$

The differential section of Higgs Boson ( $H^0$ ) of the two-photon interaction of ultraperipheral proton-proton collision can be expressed as

$$\frac{d\sigma_{pp \to pH^0p}}{d^2p_T dy} = \frac{\alpha^2}{\pi^3} 8\pi^2 (2J+1) \frac{\Gamma_{H^0 \to \gamma\gamma}}{M_{H^0}^3}$$

$$\times \frac{P_T^2}{\left(P_T^2 + \frac{\omega_1^2}{\gamma^2}\right)^2} [F_P\left(P_T^2 + \frac{\omega_1^2}{\gamma^2}\right)]^2 \int d^2 q_{2T} q_{2T}^2 \frac{[F_P(q_{2T}^2 + \frac{\omega_2^2}{\gamma^2})]^2}{\left(q_{2T}^2 + \frac{\omega_2^2}{\gamma^2}\right)^2}, \tag{10}$$

In the exclusive process, the invariant mass of the final state needs to be considered. Therefore, the differential cross section of the Higgs boson produced by the two-photon interaction in the exclusive process is give by

$$d\sigma_{H^{\pm}} = \hat{\sigma}_{\gamma\gamma \to H^+H^-}(M) dN_1(\omega_1, q_1^2) dN_2(\omega_2, q_2^2)$$

$$= d\omega_1 d\omega_2 \hat{\sigma}_{\gamma\gamma \to H^+H^-}(M) \frac{dN_1(\omega_1, q_1^2)}{d\omega_1} \frac{dN_2(\omega_2, q_2^2)}{d\omega_2},$$

where M is the invariant mass of the final state, $M_{H^{\pm}} = 155$GeV. And it is known that the equation of $\hat{\sigma}_{\gamma\gamma \to H^+H^-}(M)$ is give by

$$\hat{\sigma}_{\gamma\gamma \to H^+H^-}(M) = \frac{2\pi\alpha^2}{M^2}[(1+y)\sqrt{1-y} - 2y(1-\frac{1}{2}y)\ln(\frac{1}{\sqrt{y}} + \sqrt{\frac{1}{y}-1})], \tag{11}$$

where $y = 4M_H^2/M^2$ , $W^2 = M^2 = 4\omega_1\omega_2$ .

Therefore, the differential section of the Higgs boson ($H^{\pm}$) of ultraperipheral nuclear-nuclear collision two-photon interaction in the process of exclusive process can be expressed as[27-30]

$$\frac{d\sigma_{AB \to AH^+H^-B}}{dMd^2P_T dy} = \frac{Z_1^2 Z_2^2 \alpha^2}{\pi^2} \frac{2P_T^2}{M} \hat{\sigma}_{\gamma\gamma \to H^+H^-}(M) \times \frac{[F_N(P_T^2 + \frac{\omega_1^2}{\gamma^2})]^2}{\left(P_T^2 + \frac{\omega_1^2}{\gamma^2}\right)^2} \int d^2 q_{2T} q_{2T}^2 \frac{[F_N(q_{2T}^2 + \frac{\omega_2^2}{\gamma^2})]^2}{\left(q_{2T}^2 + \frac{\omega_2^2}{\gamma^2}\right)^2},$$

(12)

The differential section of the Higgs boson ($H^{\pm}$) of ultraperipheral proton-nuclear collision two-photon interaction in the process of exclusive process can be expressed as

$$\frac{d\sigma_{pA \to pH^+H^-A}}{dMd^2P_T dy} = \frac{Z^2 \alpha^2}{\pi^2} \frac{2P_T^2}{M} \hat{\sigma}_{\gamma\gamma \to H^+H^-}(M) \times \left\{ \frac{\left[F_N\left(P_T^2 + \frac{\omega_1^2}{\gamma^2}\right)\right]^2}{\left(P_T^2 + \frac{\omega_1^2}{\gamma^2}\right)^2} \int d^2 q_{2T} q_{2T}^2 \frac{[F_P(q_{2T}^2 + \frac{\omega_2^2}{\gamma^2})]^2}{\left(q_{2T}^2 + \frac{\omega_2^2}{\gamma^2}\right)^2} + \frac{\left[F_P\left(P_T^2 + \frac{\omega_1^2}{\gamma^2}\right)\right]^2}{\left(P_T^2 + \frac{\omega_1^2}{\gamma^2}\right)^2} \int d^2 q_{2T} q_{2T}^2 \frac{[F_N(q_{2T}^2 + \frac{\omega_2^2}{\gamma^2})]^2}{\left(q_{2T}^2 + \frac{\omega_2^2}{\gamma^2}\right)^2} \right\},$$

(13)

The cross section of Higgs boson ($H^{\pm}$) of ultraperipheral proton-proton collision two-photon interaction in the process of exclusive process can be expressed as

$$\frac{d\sigma_{pp \to pH^+H^-p}}{dMd^2P_T dy} = \frac{\alpha^2}{\pi^2} \frac{2P_T^2}{M} \hat{\sigma}_{\gamma\gamma \to H^+H^-}(M) \times \frac{[F_P(P_T^2 + \frac{\omega_1^2}{\gamma^2})]^2}{\left(P_T^2 + \frac{\omega_1^2}{\gamma^2}\right)^2} \int d^2 q_{2T} q_{2T}^2 \frac{[F_P(q_{2T}^2 + \frac{\omega_2^2}{\gamma^2})]^2}{\left(q_{2T}^2 + \frac{\omega_2^2}{\gamma^2}\right)^2},$$ (14)

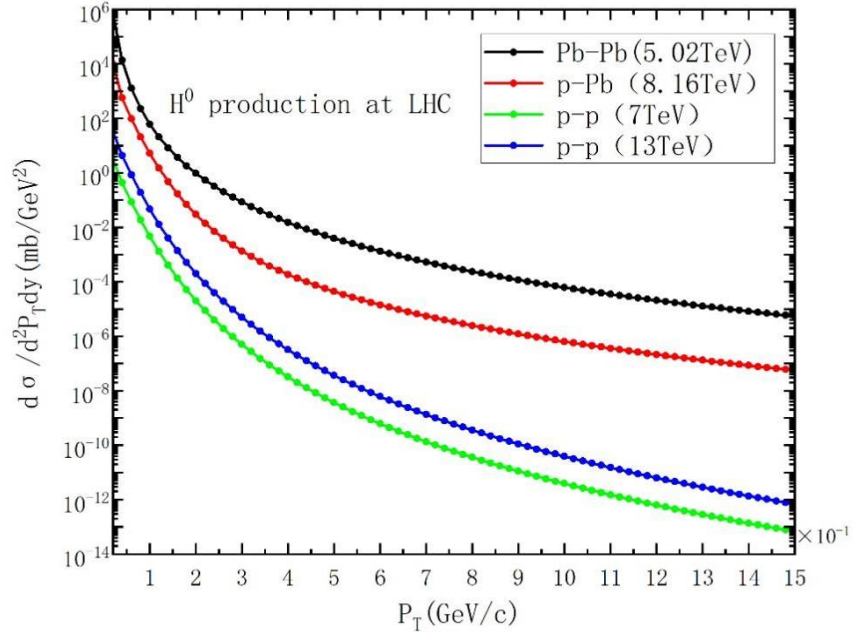

FIG. 1. The differential cross-section of the Higgs boson ($H^0$) production from the two-photon interactions in ultraperipheral heavy ion collisions at LHC. The black line is for Pb+Pb collisions with $\sqrt{s_{NN}} = 5.02$ TeV, The red line is for p+Pb collisions with $\sqrt{s_{NN}} = 8.16$ TeV, The blue line is for p+p collisions with $\sqrt{s_{NN}} = 7.0$ TeV, The green line is for p+p collisions with $\sqrt{s_{NN}} = 13.0$ TeV.

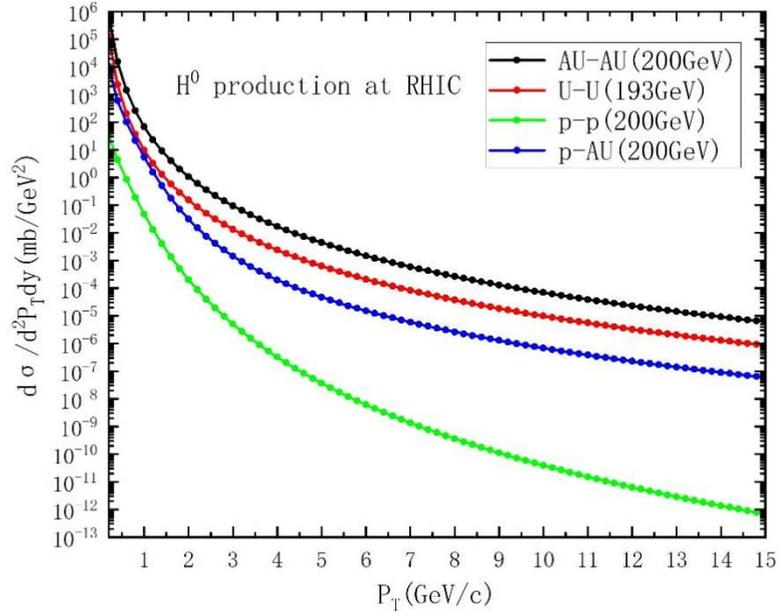

FIG. 2. The differential cross-section of the Higgs boson ($H^0$) production from the two-photon interactions in ultraperipheral heavy ion collisions at RHIC. The black line is for AU+AU collisions with $\sqrt{s_{NN}} = 200$GeV, The red line is for U+U collisions with $\sqrt{s_{NN}} = 193$GeV, The blue line is for p+AU collisions with $\sqrt{s_{NN}} = 200$GeV, The green line is for p+p collisions with $\sqrt{s_{NN}} = 200$GeV.

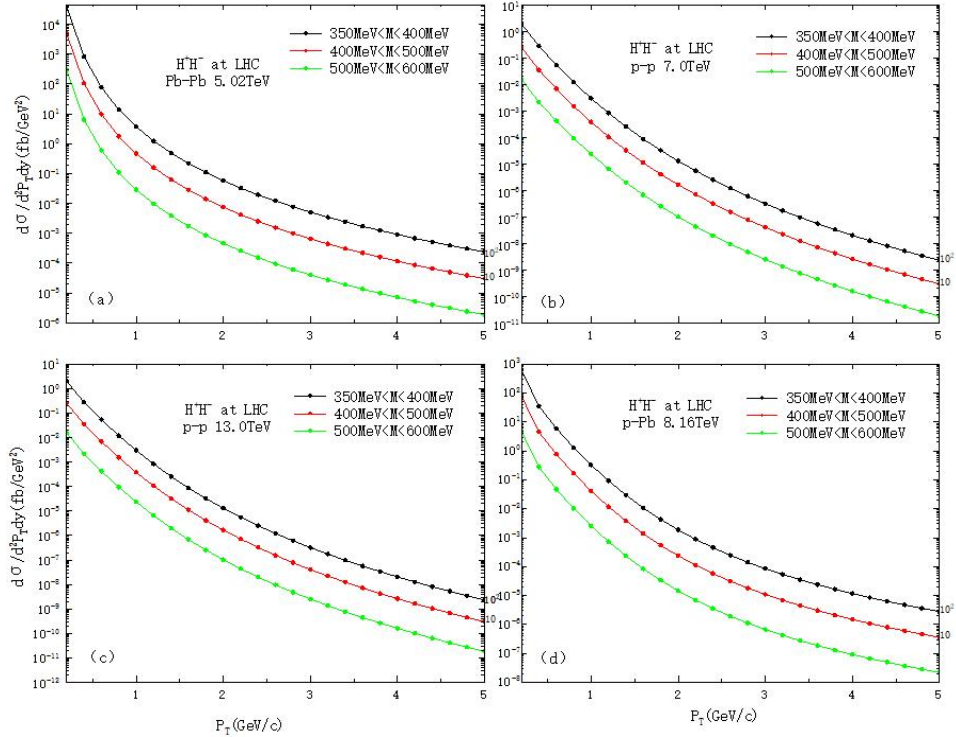

FIG. 3. The differential cross-section of the Higgs boson ($H^\pm$) production from the two-photon interactions in ultraperipheral heavy ion collisions at LHC. FIG. 3.(a) is for Pb+Pb collisions with $\sqrt{s_{NN}} = 5.02$TeV, FIG. 3.(b) is for p+p collisions with $\sqrt{s_{NN}} = 7.0$TeV, FIG. 3.(c) is for p+p

collisions with $\sqrt{s_{NN}} = 13.0$TeV, FIG. 3.(d) is for p+Pb collisions with $\sqrt{s_{NN}} = 8.16$TeV. The black line is for $350MeV < M < 400MeV$, The red line is for $400MeV < M < 500MeV$, The green line is for $500MeV < M < 600MeV$.

### III. NUMERICAL RESULTS

The differential cross section of the Higgs boson ($H^0$) produced by the two-photon interaction in the ultraperipheral heavy ion collision is positively correlated with the charge number $Z^4$ of the heavy ion, and inversely correlated with the total transverse momentum of the photon.

The differential cross section of the Higgs boson($H^0$) for large-$p_T$ production at LHC energies is potted in Fig. 1. The differential cross section of the Higgs boson($H^0$) for large-$p_T$ production at RHIC energies is potted in Fig. 2. The differential cross section of the Higgs boson($H^\pm$) for large-$p_T$ production at LHC energies is potted in Fig. 3. The differential cross section of the Higgs boson ($H^\pm$) for large-$p_T$ production at RHIC energies is potted in Fig. 4. The main sources of changes in the differential cross sections are the magnitude of the transverse momentum of the photon and the mass of the colliding nucleon, since the mass difference between heavy ions and protons is very large.

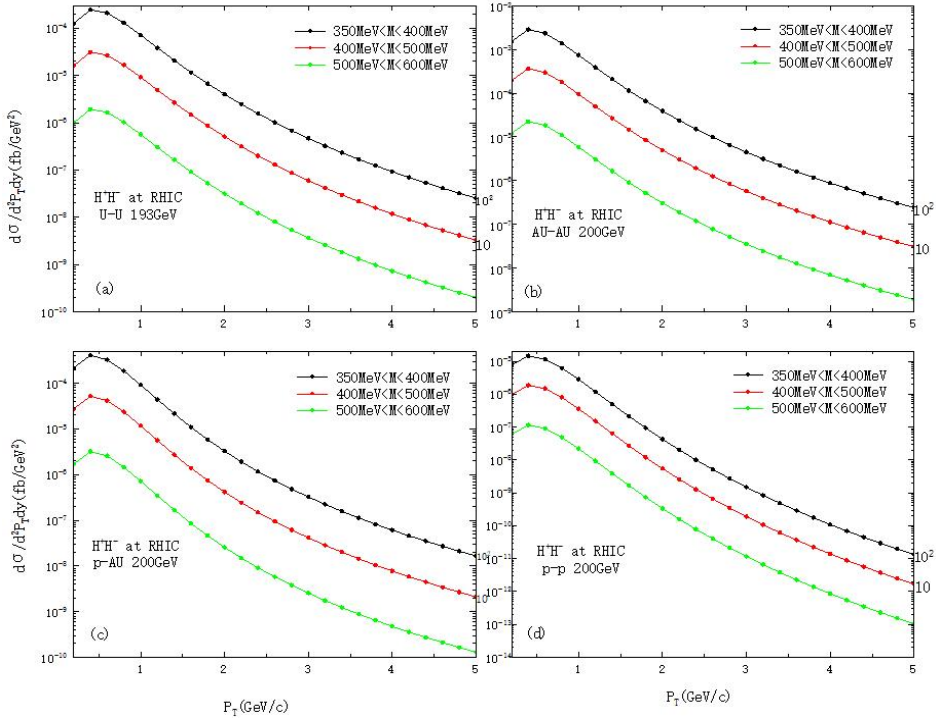

FIG. 4. The differential cross-section of the Higgs boson ($H^\pm$) production from the two-photon interactions in ultraperipheral heavy ion collisions at RHIC. FIG. 4.(a) is for U+U collisions with $\sqrt{s_{NN}} = 193$GeV, FIG. 4.(b) is for AU+AU collisions with $\sqrt{s_{NN}} = 200$GeV, FIG. 4.(c) is for p+AU collisions with $\sqrt{s_{NN}} = 200$GeV, FIG. 4.(d) is for p+p collisions with $\sqrt{s_{NN}} = 200$GeV. The black line is for $350MeV < M < 400MeV$, The red line is for $400MeV < M < 500MeV$, The green line is for $500MeV < M < 600MeV$.

## IV. CONCLUSION

In summary, we have studied the inclusive and exclusive processes, the photoproduction of ultra-peripheral heavy ion collisions with Higgs bosons at LHC and RHIC energies. In the equivalent photon approximation, the effect of the electromagnetic field for the ultra-relativistic nucleus is replaced by the flux of photons. By using a charge distribution form factor, we show the distribution of the differential cross-section and transverse momentum Higgs boson produced by semi-coherent two-photon interactions, and show the data plots at LHC and RHIC energies.

## V. ACKNOWLEDGEMENTS

This work is supported by Heilongjiang Science Foundation Project under grant No. LH2021A009, National Natural Science Foundation of China under grant No. 12063006, and Special Basic Cooperative Research Programs of Yunnan Provincial Undergraduate Universities Association under grant No. 202101BA070001-144. the National Natural Science Foundation of China under grantGrant No. 12165010, the Yunnan Province Applied Basic Research Project under grant No. 202101AT070145, the Xingdian Talent Support Project, the Young Top-notch Talent of Kunming, and the Program for Frontier Research Team of Kunming University 2023.

with two-photon processes. Chin. Phys. C 35, 109 (2011).